\documentclass[aip,graphicx,reprint]{revtex4-1}
\pdfoutput=1

\usepackage{graphicx}
\usepackage{lipsum}
\usepackage{dcolumn}
\usepackage{bm}
\usepackage[utf8]{inputenc}
\usepackage[T1]{fontenc}
\usepackage{mathptmx}
\usepackage{xcolor}
\usepackage{textgreek}
\usepackage{natbib}

\begin{document}

\title{Electrical and acoustic self-oscillations in an epitaxial oxide for neuromorphic applications.} 

\author{M. Salverda}
\email[]{m.salverda@rug.nl}
\affiliation{Zernike Institute for Advanced Materials, University of Groningen, 9747 AG Groningen, The Netherlands}

\author{B. Noheda}
\email[]{b.noheda@rug.nl}
\affiliation{Zernike Institute for Advanced Materials, University of Groningen, 9747 AG Groningen, The Netherlands}
\affiliation{CogniGron center, University of Groningen, 9747 AG Groningen, The Netherlands}

\date{\today}

\begin{abstract}
Developing materials that can lead to compact versions of artificial neurons (neuristors) and synapses (memristors) is the main aspiration of the nascent neuromorphic materials research field. Oscillating circuits are interesting as neuristors, emulating the firing of action potentials. We present room-temperature self-oscillating devices fabricated from epitaxial thin films of semiconducting TbMnO$_{3}$. We show that these electrical oscillations induce concomitant mechanical oscillations that produce audible sound waves, offering an additional degree of freedom to interface with other devices. The intrinsic nature of the mechanism governing the oscillations gives rise to a high degree of control and repeatability. Obtaining such properties in an epitaxial perovskite oxide, opens the way towards combining self-oscillating properties with those of other piezoelectric, ferroelectric, or magnetic perovskite oxides to achieve hybrid neuristor-memristor functionality in compact heterostuctures.
\end{abstract}

\pacs{}

\maketitle 

\section{Introduction}
\begin{figure*}
\includegraphics[width=0.99\textwidth]{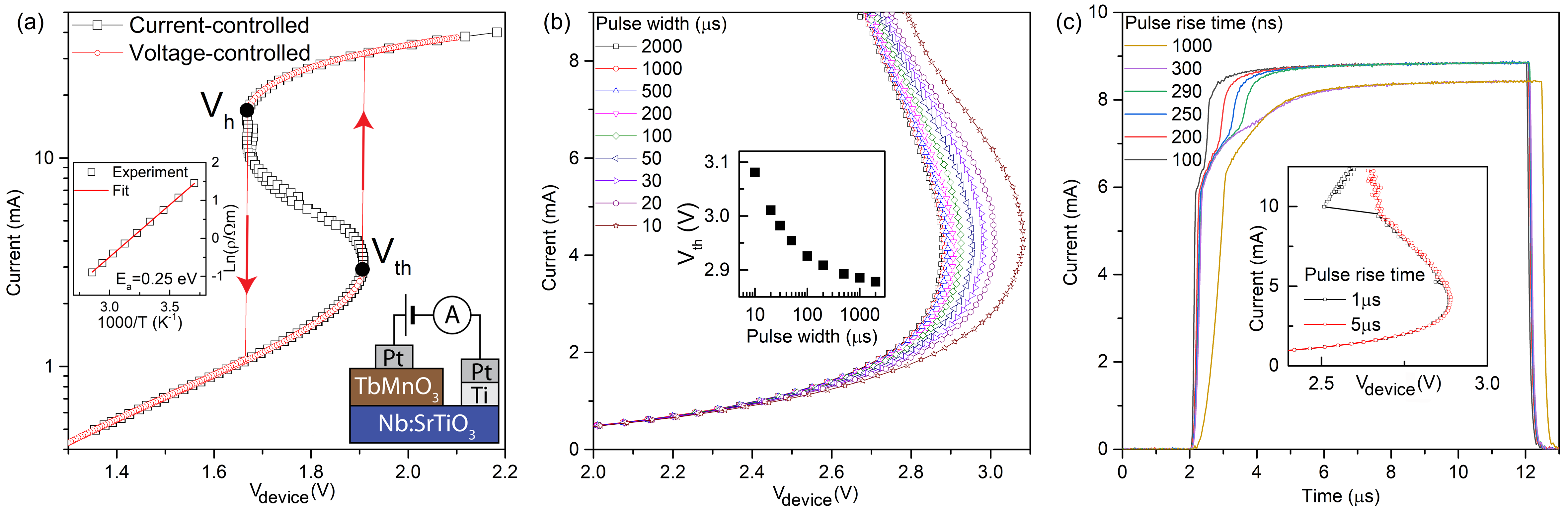}
\caption{\label{fig:Figure1} (a) Room temperature DC I-V measurement of a typical TbMnO$_{3}$ device. The V-controlled curve (red) shows the abrupt jump from the high-resistance (HR) state to the low-resistance (LR) state at V$_{th}$ and \textit{vice versa} on decreasing V for V< V$_{h}$. The current-controlled curve (black) shows a smooth transition from the HR to the LR state. Inset graph shows the thermal activation of the transport and inset cartoon shows the schematic device structure; (b) Pulsed I-V measurement for different pulse widths ($\Delta t$).  V$_{\text{device}}$ was calculated using the I, applied V and a series resistance R$_s$= 500 \textOmega; Inset shows the V$_{th}$ values for different $\Delta t$;  (c) Time-resolved I measurement during pulses with V= 7.45 V,  $\Delta t$= 10 \textmu s, R$_s$= 600 $\Omega$ and varying rise times (t$_r$). Below a certain t$_r$, a transition is observed to a resistance state than is even lower than that observed for the higher t$_r$'s. Inset: Pulsed I-V measurement with different t$_r$'s ($\Delta t$= 1 ms).}
\end{figure*}

Current-controlled Negative Differential Resistance (CC-NDR) occurs when the voltage (V) across a material decreases while the current (I) increases. I-V characteristics featuring CC-NDR display a region of voltages, between the threshold voltage, V$_{th}$, and the holding voltage, V$_{h}$, where multiple current values are possible for a single voltage (see figure \ref{fig:Figure1}(a)). CC-NDR can be caused by different mechanisms, such as impact ionization\cite{Mayaram1993,Berglund1971,Altcheh1973} and metal-insulator transitions\cite{Zhou2015,Pickett2012b,Valmianski2018,Zhou2015,Imada1998}, but also by thermally activated and assisted mechanisms, such as Schottky emission and Poole-Frenkel conduction.\cite{Gibson2018,Funck2016,Gibson2016}. Although material systems with CC-NDR have been found since the 60's\cite{Kaplan1972,Knight1972,Haubenreisser1974,Fischer2013,Biskup2006,Arya1979,Cope1965,RoyBardhan1974}, the promise of energy-efficient, neuromorphic hardware to perform in-memory and on-edge processing\cite{Ielmini2020,Burr2017,Sebastian2020,Chua2012a,Chicca2020,DelValle2018}, has renewed interest and brought about a new wave of research on this phenomenon.\cite{Goodwill2019a,Kumar2017a,Gao2017,Zhou2015,Andrews2019,Rozenberg2019} For instance, selector devices are being implemented in memristor arrays, where the highly non-linear NDR helps to reduce leakage and sneak path currents.\cite{Burr2014} CC-NDR is also used to construct self-oscillators with regular and chaotic oscillations and to emulate spiking neurons.\cite{Yi2018,DelValle2018,Zhou2015,Pickett2013,Ascoli2015,Herzig2019,Andrews2019,Rozenberg2019}

Although most recent work is focused on transition metal-oxide systems, such as TaO$_{x}$\cite{Goodwill2018,Gala2016,Goodwill2019a}, VO$_{x}$\cite{Zhou2015,Mian2015} and NbO$_{x}$\cite{Zhou2015,Herzig2019,Li2015}, the latter two also displaying Mott transitions above room temperature, another class of well-known materials are oxide perovskites. Heterostructures of these perovskites are rich in functional properties\cite{Coll2019} and memristive and synaptic devices have been fabricated with them.\cite{Chanthbouala2012,Goossens2018,Bagdzevicius2017,Acevedo2016,DelValle2018,Guo2020} Selectors and neuron-like devices made of these materials could allow for heterostructures that combine neuristor and synaptic functionality in a compact manner. However, considering the many mechanisms that can lead to CC-NDR, only relatively few perovskite oxides have been shown to exhibit it, and most of these studies were performed at low temperatures.\cite{Asamitsu1997,Biskup2006,Lu2006,Jain2007}

Here we present room-temperature CC-NDR in devices fabricated from epitaxial thin films of semiconducting TbMnO$_{3}$. We show that the device can oscillate between high and low resistance states and that these electrical and thermal oscillations also give rise to mechanical oscillations producing detectable and audible sound waves with extraordinary endurance. 

\section*{Results}
I-V measurements were carried out on TbMnO$_3$ thin films under DC voltage, which display CC-NDR in a highly reproducible manner without the need of a forming step. Typical CC-NDR behaviour is observed with a smooth transition for current-controlled I-V measurements (also referred to as S-type NDR), while  voltage-controlled measurements show hysteresis (see figure \ref{fig:Figure1}(a)). At low V (< 1 V), an exponential I \textit{vs.} V dependence is found (see Supplementary Information), originating from the p-n junction at the interface between the p-type TbMnO$_{3}$ and n-type Nb-doped SrTiO$_{3}$.\cite{Jin2014,Cui2013} At higher voltages (1 V < V < 1.68 V), the current is limited by series resistances, including that of the TbMnO$_{3}$ film and the Nb-doped SrTiO${3}$ substrate, and becomes linear. Above 1.68 V, the transport becomes nonlinear again, ultimately leading to the NDR regime with V$_{th}$= 1.9 V (ON-switching). On the return sweep, the OFF-switching takes place at V$_{h}$= 1.68 V. The horizontal axes display the device voltage, V$_{device}$, corrected for a series resistor added to prevent current overshoots, instead of the applied voltage, V. It should also be noted that a significant voltage will drop over the p-n junction at the interface (up to 0.9 V), which means that the voltages that are actually applied to the TbMnO$_{3}$ film are smaller than what is indicated by the horizontal axes in the figures. 

A measurement of the resistance of the film shows that the electrical transport of TbMnO$_{3}$ around and above room temperature is thermally activated, following 
\begin{equation}
    \sigma = \sigma _{0} e^{-E_{a}/k_{B}T}
    \label{eq:ActivatedTransport}
\end{equation}
with an activation energy E$_{a}$ of about 0.25 eV (see inset in figure \ref{fig:Figure1}(a)), which is consistent with an activation energy across a semiconducting band gap of E$_{g}$ = 2E$_{a}$ = 0.5 eV, in agreement with the reported values in this material.\cite{Cui2005} Current passing through the device causes Joule heating, which increases the temperature of the device as $ T=T_{amb}+R_{th}IV,$ where R$_{th}$ is the thermal resistance of the device, which includes the heat capacity and thermal conductance of the active region to the surrounding material. Together with eq. \ref{eq:ActivatedTransport}, this  gives rise to the CC-NDR behaviour: the dissipated power increases the device temperature, which leads to a decrease in its resistance.\cite{Muller1976,Altcheh1973,Brodkorb1972,Brodkorb1972a} This is not only true for the activated electrical transport of eq. \ref{eq:ActivatedTransport}, but for all conduction mechanisms that depend superlinearly on temperature\cite{Gibson2018,Altcheh1973,Berglund1971} There are also mechanisms that are field-dependent, such as avalanche/impact ionization\cite{Mayaram1993,Altcheh1973,Berglund1971}, often found in semiconductors as well. However, as the current usually increases enormously when these mechanisms are triggered, producing additional Joule heating, they are often difficult to separate from temperature-activated effects.\cite{Altcheh1973,Berglund1971} Another well-known mechanism is the Mott metal-insulator transition\cite{Valmianski2018,Zhou2015,Imada1998, Pickett2012b}, which can be triggered by temperature or electric field. TbMnO$_{3}$ does not have such a transition.

\subsection*{Pulsed I-V and transient measurements}
 A temperature effect requires some settling time, while an electric field effect would show itself much faster (provided a small electrical RC\cite{Goodwill2019,Wang2018b}). Pulsed I-V measurements can, therefore, attest that thermal processes are at play. Figure \ref{fig:Figure1}(b) shows the measurements for different pulse lengths, from $\Delta t$= 10 \textmu s to 2 ms. The inset shows the threshold voltage V$_{th}$ obtained from these measurements: shorter pulses drive V$_{th}$ to higher values, which is a result of the temperature not reaching its steady-state value. For longer pulse widths, V$_{th}$ approaches a constant value, which indicates that the device temperature is close to its steady state value corresponding to that applied voltage.

Interestingly, at higher currents within the NDR region, we observe that the device can further decrease its resistance if the rise time (t$_r$) of the pulse is short, as shown in the inset of figure \ref{fig:Figure1}(c) for t$_r$= 1 \textmu s compared to t$_r$= 5 \textmu s). To investigate the origin of this phenomenon, we perform time-resolved current measurements during the application of voltage pulses with various  t$_r$ (see figure \ref{fig:Figure1}(c)). These results shows that a smooth transition to a higher current state takes place for t$_r$< 300 ns.

\begin{figure*}
\includegraphics[width=0.99\textwidth]{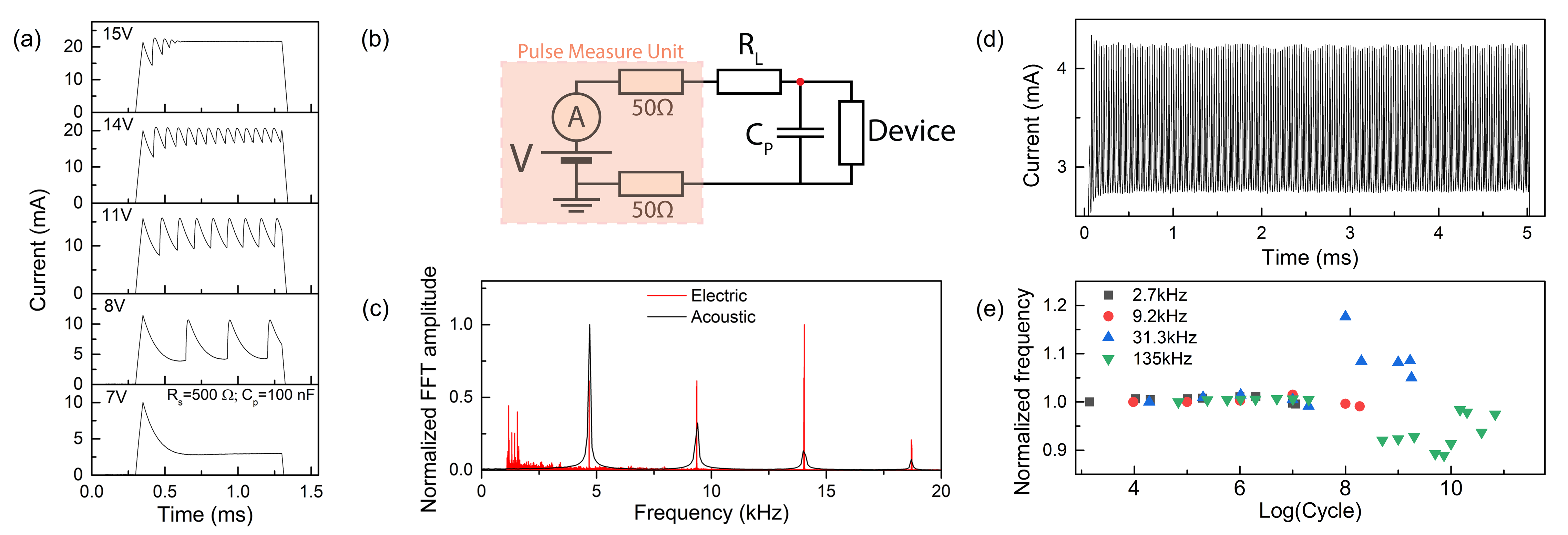}
\caption{\label{fig:Figure2} (a) Time-resolved I measurements showing oscillations when 7V < V< 15 V, with V being the applied voltage (different from the V$_{device}$ ;(b) Schematic circuit used to measure and manipulate the oscillations. The load resistance R$_{L}$ and the parallel capacitance C$_{P}$ are varied using a decade box, while another source of resistance comes from the Pulsed Measure Unit;(c) Comparison of the Fourier transform of the electric oscillations, measured by the oscilloscope, and the acoustic signal, measured by the microphone;(d) Example of sustained oscillations without damping;(e) Endurance measurement in four fresh devices, using four different values of parallel capacitance.}
\end{figure*}

The observation of this additional transition can be explained by the formation of a high-current density domain, induced by the local increase of temperature, which has also been referred to as "thermal filament".\cite{Kumar2018,Wang2018b,Goodwill2019,Goodwill2017,Haubenreisser1974,Li2017,Nath2019,Goodwill2019a,Nandi2019} If the electrical relaxation time of the measurement circuit, which is determined by the resistance and capacitance of the circuit and the device, is slower than the thermal relaxation time, the temperature of the device can keep up with the electrical input and the device heats up gradually and relatively uniformly, as if it were in a semi steady-state during the increase of voltage. As the temperature of the material increases, its resistance goes down and a larger fraction of the applied voltage will drop in the series resistance (if there is one), limiting the electric field over the material. If, on the other hand, the electrical relaxation time is faster than the thermal one, the electric field on the material increases faster than the temperature can keep up. A higher electric field than in the previous case is reached across the device. In this scenario, field-activated transport, such as Poole-Frenkel conduction, which is sensitive to factors like roughness and inhomogeneities in the material, can be induced. Locally, large currents flow and high temperatures develop.\cite{Goodwill2019,Kumar2018,Nath2019,Nandi2019} 

\subsection*{Electrical oscillations and sound}
As mentioned in the introduction, CC-NDR can be used to make oscillators. To do this in our system, we place a resistor in series and a capacitor in parallel to the device (see figure \ref{fig:Figure2}(b)). This circuit is much like that of the Pearson-Anson oscillator\cite{Pearson1921}, a type of Van der Pol oscillator.\cite{VanderPol1926} Under and applied voltage (V), V$_{device}$ will increase gradually, until V$_{th}$ is reached. The device then switches, its resistance dropping faster than the voltage can keep up (caused by the larger electrical RC), and the charge that was stored on the capacitor is discharged through the device. As a result of the discharge, the voltage over the device drops below V$_{h}$, after which the current decreases abruptly, allowing the temperature of the device to decrease and returning the device to its HR state. Subsequently, the capacitor is allowed to charge again, which marks the beginning of the next oscillation. 

Figure \ref{fig:Figure2}(a) shows a series of measurements for various pulse voltages, in which we observe the self-oscillations. A pulse with a voltage that induces V$_{device}$< V$_{th}$ does not lead to oscillations (7V in this example). For V$\geq$ 15 V, the frequency and total power become too large for the device to be able to cool down in between two heating events. In this condition, the oscillations are damped and the system ends up in the high-temperature, LR state. Note that the voltages are rather high, which is the result of the large series resistance we use (500 $\Omega$) in order to observe the oscillations.  Beside the applied voltage, two other parameters that play a role in the occurrence and frequency of oscillations in such a circuit are the load resistance and parallel capacitance\cite{Liu2016a,Lalevic1981} (see also Supplementary Information, figures S2 and S3). 

For frequencies below 20 kHz, the oscillations produce audible sound waves. We recorded the audio signal with a microphone, with the spectrum of one such a recording shown in figure \ref{fig:Figure2}(c), upon application of a DC voltage using a Source Measure Unit (SMU). Since we are sourcing a DC voltage, rather than using the Pulse Measure Unit (PMU), we use an oscilloscope to record the electrical signal, which we connect to the node in between the device and the series resistance (see the red node in figure \ref{fig:Figure2}(b)). The Fourier transform of the electrical signal is overlaid with the acoustic spectrum in figure \ref{fig:Figure2}(c), clearly showing that the electrical and mechanical oscillations have the same origin. A possible explanation for this observation is the thermal expansion and contraction induced in the material. 

This concomitant mechanical response can raise questions about the robustness of the oscillations and the possible aging of the material. Therefore, we have investigated the endurance of these devices. We use the same setup as mentioned in the paragraph above, only this time we vary the parallel capacitor in order to vary the oscillation frequency. The results for four different devices with four different oscillation frequencies are shown in figure \ref{fig:Figure2}(e) and an example of sustained oscillations in figure \ref{fig:Figure2}(d). We find that the oscillator endures at least 10$^{11}$ cycles at 135 kHz, while the devices set to oscillate at lower frequencies are stable for at least 10$^{7}$ cycles. These values act as a lower limit for the actual endurance, as the measurement were deliberately stopped after a few hours (or days, in the case of 135 kHz). The observed variability above 10$^{8}$ cycles could be due to a slow change in the ambient conditions over the course of hours. We also observed that some devices would stop oscillating occasionally, but adjusting the voltage was enough to re-initiate the oscillations. Probably a change in ambient conditions caused the device to drop out of the parameter range for the devices to oscillate. We have not seen any device breaking down during or after cycling. As we varied the frequency by varying the parallel capacitance, the lower frequency oscillations are accompanied by larger discharge currents when the device switches. The fact that these devices can withstand these currents (and the accompanying temperatures, which can reach several hundreds of Kelvin higher locally\cite{Goodwill2019,Kumar2018,Nath2019,Nandi2019}) attest to the robustness of the devices.

\section*{Discussion}

Several remarks can be made about these results. Firstly, for these devices to be used as selectors in memristor arrays, ideally V$_{th}$< 1 V, the ON-OFF ratio larger than 1000 and the current in the OFF-state as low as possible\cite{Li2017b,Wang2019}, these requirements likely being similar for efficient oscillators. These characteristics are less favourable in the devices described in this work than in devices made of NbO$_{2}$, VO$_{2}$ and TaO$_{x}$ thus far, as expected, as these other systems have been optimized for this purpose for many years. The expectation is that the optimization in TbMnO$_3$ can be accelerated as some of the insights from that work, such as the role of device geometry and dimensions, can directly be transferred to this material.\cite{Wang2018b,Liu2016a,Liu2014,Cha2016}

Additional support for this expectation is that in NbO$_{2}$ and TaO$_{x}$ the mechanism causing the CC-NDR is believed to be the Poole-Frenkel effect. In the low-field regime, NbO$_{2}$ has an activation energy in the range of 0.3-0.67 eV\cite{Wang2018b,Wang2018a,Funck2016} and TaO$_{x}$ in the range of 0.24 eV-0.44 eV\cite{Goodwill2017}, both depending on oxygen content,  This compares well with that of TbMnO$_{3}$. Although it is not clear to what extent the Poole-Frenkel mechanism plays a role in TbMnO$_{3}$, it is likely that it does, as TbMnO$_{3}$ has a smaller dielectric constant than these two materials.\cite{Goodwill2017,Cui2011,Funck2016}



Moreover, the large variety of polymorphs in transition metal simple oxides such as NbO$_{2}$ and TaO$_{x}$ is a concern, as their electric properties (especially the activation energy) depend strongly on their stoichiometry\cite{Goodwill2017,Wang2018a,Funck2016,Jeong2013} In contrast, in preparation for this work we have deposited TbMnO$_{3}$ under various growth conditions, yet the resulting activation energy is nearly identical ($\approx$ 0.25 eV). See for more details table S1 in the Supplementary Information. 

Several reports on NbO$_{2}$\cite{Nandi2017,Liu2016b,Ascoli2015} and TaO$_{x}$\cite{Gala2016} indicate that a forming step is required to initiate threshold switching and fast oscillations in large devices. Forming confines the current in smaller volume filaments that can heat up and cool down faster, producing faster oscillation frequencies.\cite{Liu2014,Funck2017,Wang2019}. In this work we did not require a forming step; however, we did observe that, whenever we deliberately damage a device by passing a high current through it ($>$100 mA) the oscillation frequency would increase substantially. The damage is observed as a permanent change in the electrical behaviour of the device as well as a visually damaged electrode. The increase in frequency could be explained either by the formation of a filament with different chemical composition or by the delamination of the electrode caused by the extreme heat that is produced during the thermal runaway\cite{Goodwill2019,Dittmann2012,Munstermann2010}. Both would lead to a smaller volume that is heated and therefore a higher oscillation frequency. Compared to requiring forming steps or inducing damage, miniaturization of these device is a more controlled way to obtain higher frequencies.

Additionally, as mentioned earlier, the voltages reported in this work include a voltage drop on the p-n junction at the interface between the TbMnO$_{3}$ film and the Nb-doped SrTiO$_{3}$ substrate. In order to operate the device at a lower voltage, this extra voltage drop is unwanted. Making use of a lateral device geometry, e.g. on non-doped SrTiO$_{3}$, or by making use of a bottom electrode that forms an ohmic interface with TbMnO$_{3}$ would both prevent this issue. 

Coupling oscillators is of great interest to neuromorphic computing strategies\cite{Chua2005a,Chua2012a,Yi2018,Romera2018}. The coupling can be done electrically\cite{Pickett2013,Yi2018,Romera2018}, but recently it was shown that these oscillators based on Joule heating can be coupled through temperature as well\cite{Velichko2018}. The mechanical response of the TbMnO$_{3}$ in this work potentially allows for an additional way of coupling, not only to other oscillators but also to piezoelectric materials, which can be in the same heterostructure or in a distant device, using the acoustic signal as wireless communication. 

Finally, the phenomena shown here are also expected in other perovskites with similar electronic band structures, enhancing the prospects for combinations of different neuromorphic functionalities in a single compact heterostructure. Indeed, in studies about unrelated phenomena, several perovskite manganites have shown CC-NDR, albeit in bulk form or (well) below room temperature (e.g. the well-known manganite Pr$_{1-x}$Ca$_{x}$MnO$_{3}$) possibly all caused by Joule heating.\cite{Biskup2006,Mercone2005,Fisher2006} Also, several perovskites exist that show an insulator-metal transition, such as nickelates\cite{Catalan2008a,Torrance1992,Catalano2018} and several manganites\cite{Coey1999,Salamon2001}, which would, just like in NbO$_{2}$ and VO$_{2}$, enhance the ON-OFF ratio of these devices. In addition, the transition temperature in these materials is tunable by composition\cite{Catalan2008a,Torrance1992,Catalano2018,Andrews2019,Ogimoto2005} and strain,\cite{Catalano2018,Andrews2019,Ogimoto2005} granting additional design flexibility.

\section*{Conclusion}
In summary, we have shown that CC-NDR is found in thin films of TbMnO$_{3}$. All the experiments, including pulsed I-V measurements, are consistent with Joule heating as the origin of the NDR effect. We found that the rate at which the voltage is applied has an influence on the final resistance of the device, which could indicate the formation of high-temperature current constriction. As expected, the device produces current oscillations if a Pearson-Anson circuit is created with it. The frequency of the electrical oscillations reaches up to 200 kHz, while below 20kHz the oscillations are concomitant with sound waves of the same frequency, which can be detected by a piezoelectric. Even though the sound waves indicate a large mechanical response of the material, the endurance is high. The limits of the endurance have not been tested in this work. This work could be a first step on the development of oxide perovskite heterostructures with combined neuromorphic functionalities. Future research should be performed to design devices that can profit from both the electrical and acoustic nature of the oscillations. As an example, oscillator networks could benefit from the possibility of using acoustic waves to do the coupling or low frequency electrical signals could be transmitted wireless through limited distances using a piezoelectric receiver.

\section*{Methods}

Following previous work\cite{Farokhipoor2014}, epitaxial thin films of TbMnO$_{3}$ with a thickness of 60 nm were grown on single-terminated Nb-doped SrTiO$_{3}$ substrates by Pulsed Laser Deposition, using an oxygen pressure of 0.9 mbar, a laser fluence of 2.35 J/cm$^2$, a heater temperature of 750 $^\circ$C, a target-heater distance of 55 mm and repetition rate of 1 Hz, produced by a 248 nm excimer laser. 

On top of the film, circular Pt electrodes with a thickness of 80 nm and diameter of 10 \textmu m were fabricated using photolithography and e-beam evaporation. To provide an ohmic contact with the Nb-doped SrTiO$_{3}$ electrode\cite{Park2008}, a layer of Ti (10 nm) was evaporated to the bottom side of the substrate, capped with a layer of Pt (50 nm) (figure \ref{fig:Figure1}(a)). Electrical measurements were performed using a Keithley 4200A-SCS parameter analyzer equipped with Source Measure Units (SMUs) and a Pulse Measure Unit (PMU), and a LeCroy 9410 Oscilloscope. Audio recordings were taken using a Samson C02 commercial condenser microphone and a Zoom H6 interface.

\section*{Data availability}
The data that support the findings of this study are available from the corresponding author upon reasonable request.

\section*{References}
\bibliography{ms.bib}

\section*{Acknowledgments}
We thank Cynthia Quinteros and Pavan Nukala for their discussions and contributions during this project and Arjun Joshua for technical support. Financial support by the Groningen Cognitive Systems and Materials Center (CogniGron) and the Ubbo Emmius Foundation of the University of Groningen is gratefully acknowledged.

\section*{Author contributions}
M.S. and B.N. designed the project. M.S. did the experimental work (sample growth and characterization, device fabrication and electrical measurements) and analysed the data. M.S. and B.N. discussed the results. M.S. wrote the first version of the manuscript that was reviewed by both authors.

\section*{Competing interests}
The authors declare no competing interests.

\end{document}


\title{\bf Supplementary Information: Electrical and acoustic self-oscillations in an epitaxial oxide for neuromorphic applications.}

\author{Mart Salverda}
\email{m.salverda@rug.nl}
\affiliation{Zernike Institute for Advanced Materials, University of Groningen, 9747 AG Groningen, The Netherlands }
\author{Beatriz Noheda}
\email{b.noheda@rug.nl}
\affiliation{Zernike Institute for Advanced Materials, University of Groningen, 9747 AG Groningen, The Netherlands }
\affiliation{CogniGron center, University of Groningen, 9747 AG Groningen, The Netherlands }

\maketitle
\onecolumngrid

\beginsupplement

\begin{figure*}[hbp]
	\includegraphics[width=12cm]{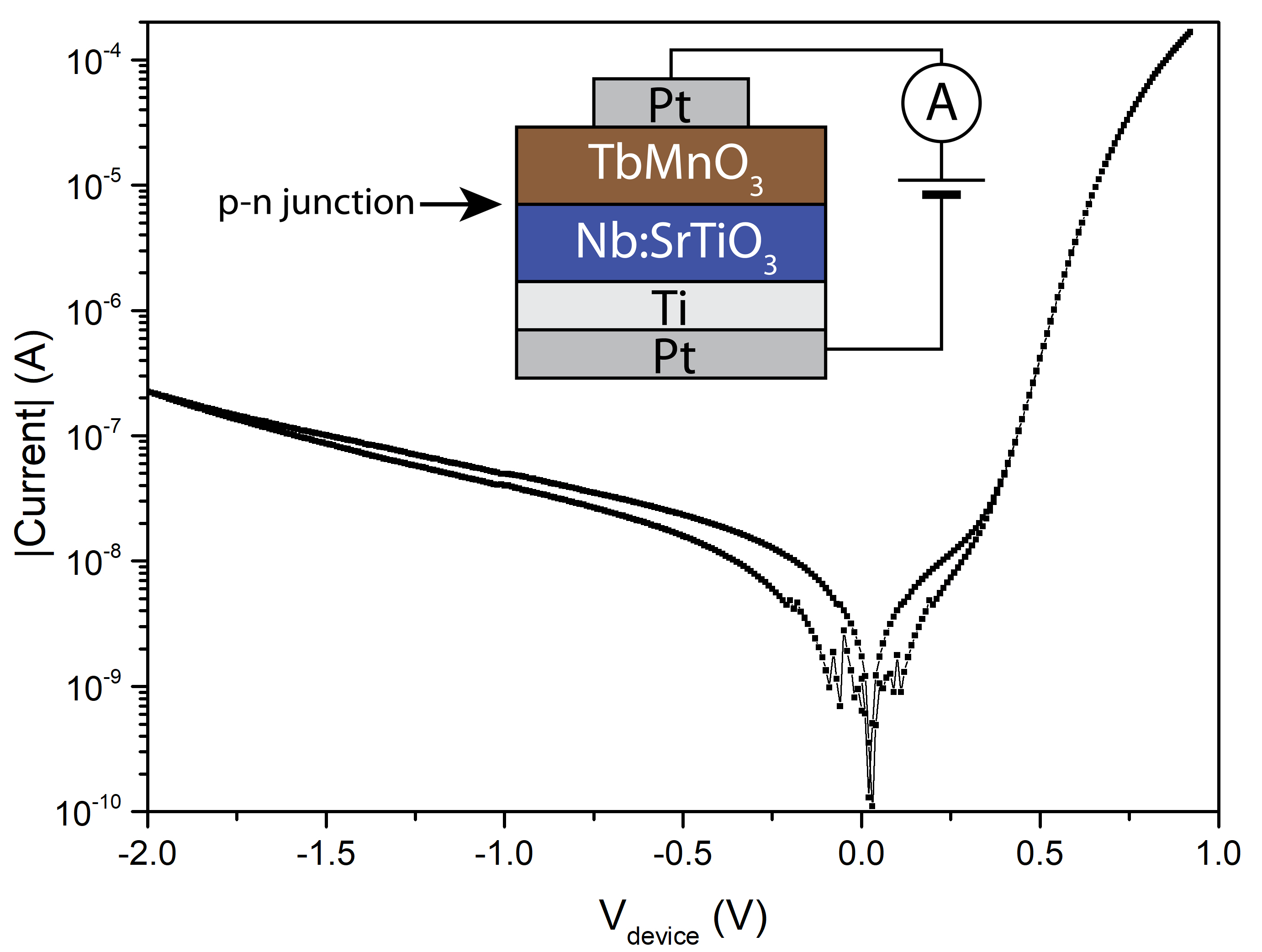}
	\centering 
	\caption{\label{fig:S-T-dependence} 
At low voltages and currents, the I-V measurement shows a clear diode characteristic, caused by the p-n junction at the interface of the TbMnO$_{3}$ film and Nb-doped SrTiO$_{3}$ substrate.
	}
\end{figure*}

\begin{figure*}[hbp]
	\includegraphics[width=12cm]{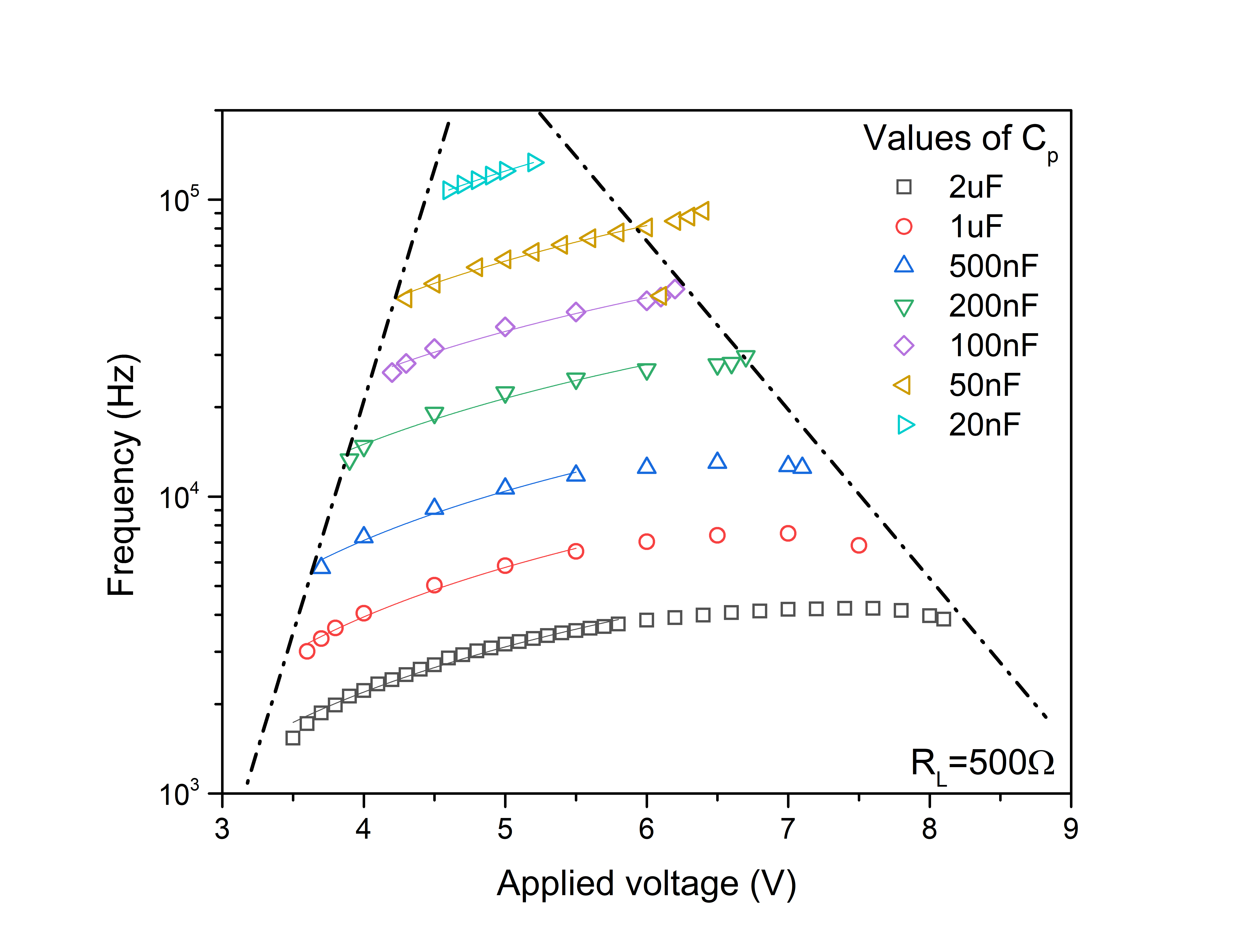}
	\centering 
	\caption{\label{fig:S-FrequencyVersusVoltage_VaryingCapacitors} 
Oscillation frequency as function of applied voltage for various parallel capacitances C$_{p}$. For each capacitance there is a maximum and a minimum voltage above and below which the device does not oscillate. Within the range, there is a smaller range in which the frequency depends linearly on voltage.
	}
\end{figure*}

\begin{figure*}[hbp]
	\includegraphics[width=12cm]{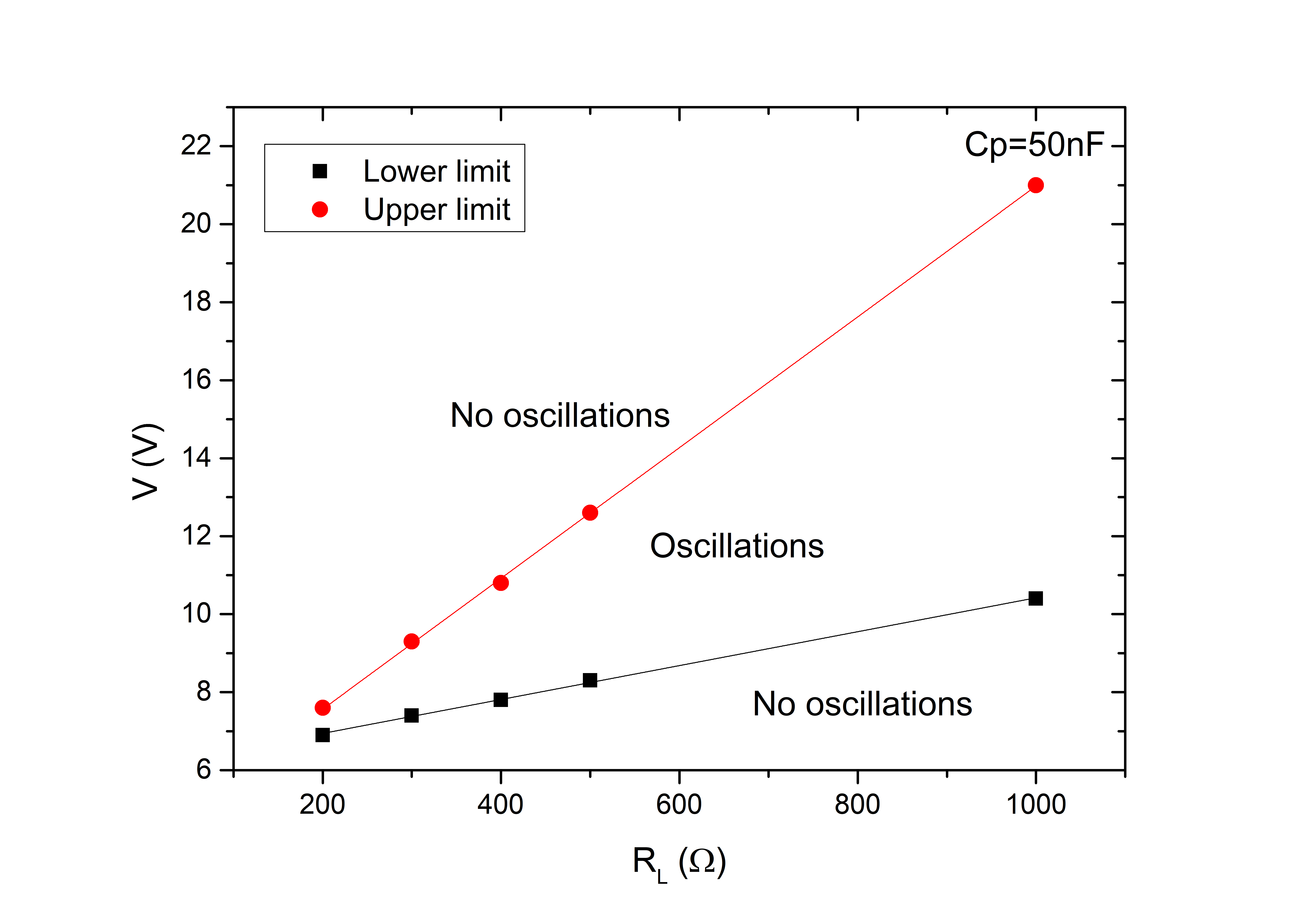}
	\centering 
	\caption{\label{fig:S-UpperLowerLimits} 
Upper and lower limits of the oscillation range for a fixed C$_p$ and varying applied voltage and R$_L$.
	}
\end{figure*}

\begin{table}[hbp]
\begin{tabular}{|l|ccccc|}
\hline
Sample name & \multicolumn{1}{c}{Pressure} & \multicolumn{1}{c}{Temperature} & \multicolumn{1}{c}{Fluence} & \multicolumn{1}{c}{Frequency} & \multicolumn{1}{c|}{E$_{a}$} \\
       & \multicolumn{1}{c}{(mbar)}         & \multicolumn{1}{c}{($^{\circ}$C)}            & \multicolumn{1}{c}{(J/cm$^{2}$)}        & \multicolumn{1}{c}{(Hz)}          & \multicolumn{1}{c|}{(eV)}   \\ \hline
MS158       & 0.5                          & 750                             & 2                           & 1                             & 0.23                    \\
MS159       & 0.25                         & 750                             & 2                           & 1                             & 0.26                    \\
MS160       & 0.5                          & 750                             & 2                           & 0.5                           & 0.27                    \\
MS185       & 0.007                        & 800                             & 1                           & 5                             & 0.28                    \\
MS191       & 0.01                         & 800                             & 1                           & 5                             & 0.26                    \\
MS193       & 0.01                         & 800                             & 2.35                        & 1                             & 0.23                    \\
MS147       & 0.9                          & 750                             & 2.35                        & 1                             & 0.24                    \\ \hline
\end{tabular}
\caption{\label{tab:S-activationenergies} Activation energies for films grown at various growth parameters. The pressure and energy are varied over a rather large range, yet the activation energy hardly differs. }
\end{table}
